\def\gpath{.} 

\documentclass[%
reprint,
superscriptaddress,
showpacs,
aps,
prl,
]{revtex4-1}

\usepackage{graphicx}

\begin{document}


\title{{Cavitating Langmuir Turbulence in the Terrestrial Aurora}}


\author{B.\ Isham}
\email[]{bisham@email.bc.inter.edu}
\affiliation{
Department of Electrical and Computer Engineering,
Interamerican University of Puerto Rico, 
Bayam{\'o}n, Puerto Rico 00957, USA
}

\author{M.~T.\ Rietveld}
\affiliation{
European Incoherent Scatter Scientific Association,
9027 Ramfjordbotn, Norway
}

\author{P.\ Guio}
\affiliation{
Department of Physics and Astronomy, 
University College London,
London, WC1E 6BT, United Kingdom
}

\author{F.~R.~E.\ Forme}
\affiliation{
Centre d'{\'E}tude Spatiale des Rayonnements, 
31028 Toulouse, France
}

\author{T.\ Grydeland}
\affiliation{
Northern Research Institute Troms{\o}, 
Postboks 6434 Forskningsparken, 9294 Troms{\o}, Norway
}

\author{E.\ Mj{\o}lhus}
\affiliation{
Department of Mathematics and Statistics, 
University of Troms{\o}, 
9037 Troms{\o}, Norway
}


\date{\today}

\begin{abstract}
Langmuir cavitons have been artificially produced in the earth's ionosphere, 
but evidence of naturally-occurring cavitation has been elusive. 
By measuring and modeling the spectra of electrostatic plasma modes, 
we show 
that natural cavitating, or strong, Langmuir turbulence 
does occur
in the ionosphere, 
via a process in which a 
beam of 
auroral 
electrons drives Langmuir waves, which 
in turn 
produce cascading Langmuir and ion-acoustic excitations 
and cavitating Langmuir turbulence.
The data presented here 
are the
first direct evidence 
of cavitating Langmuir turbulence occurring naturally 
in any 
space or 
astrophysical plasma. 
\end{abstract}


\maketitle





Langmuir turbulence is 
known to occur in controlled laboratory 
 \cite{
 WongCheung_1984_PRL,
 VyacheslavovA_2002_PlPhCF}
and space plasma experiments 
 \cite{
SulzerFejer_1994_JGR, 
IshamA_1999_PRL, 
RietveldA_2000_JGR} 
and 
is thought
to occur naturally in 
a variety of 
space and astrophysical plasmas, 
including 
pulsar magnetospheres 
 \cite{
AsseoPorzi_2006_MNRAS},
the solar corona 
\cite{
NulsenA_2007_JGR},
%
the interplanetary medium 
\cite{KelloggA_1992_GRL}, 
%
planetary foreshocks
\cite{
RobinsonCairns_1995_GRL}, 
%
the terrestrial magnetosphere 
\cite{StasiewiczA_1996_JGR}, 
and 
the ionosphere 
%
%
\cite{
PapadopoulosCoffey_1974_JGR_nonthermal, 
RowlandA_1981_PRL, 
GuioForme_2006_PP}. 
%
%
In its most developed form, this turbulence contains electron Langmuir modes trapped in dynamic density depressions known as cavitons
 \cite{
Robinson_1997_RMP, 
DuBoisA_1991_PRL, 
DuBoisA_1993_PhysFluids}. 
Cavitons have been shown to be artificially produced in the earth's ionosphere during high-power radiowave pumping experiments as deduced from radar spectra containing simultaneously-excited up and down-shifted Langmuir and ion-acoustic lines plus a central peak due to cavitation 
 \cite{
SulzerFejer_1994_JGR, 
IshamA_1999_PRL, 
RietveldA_2000_JGR}, 
but evidence of naturally occurring cavitation has until now been elusive. 

\begin{figure}[t] 
\includegraphics[scale=0.98]{\gpath/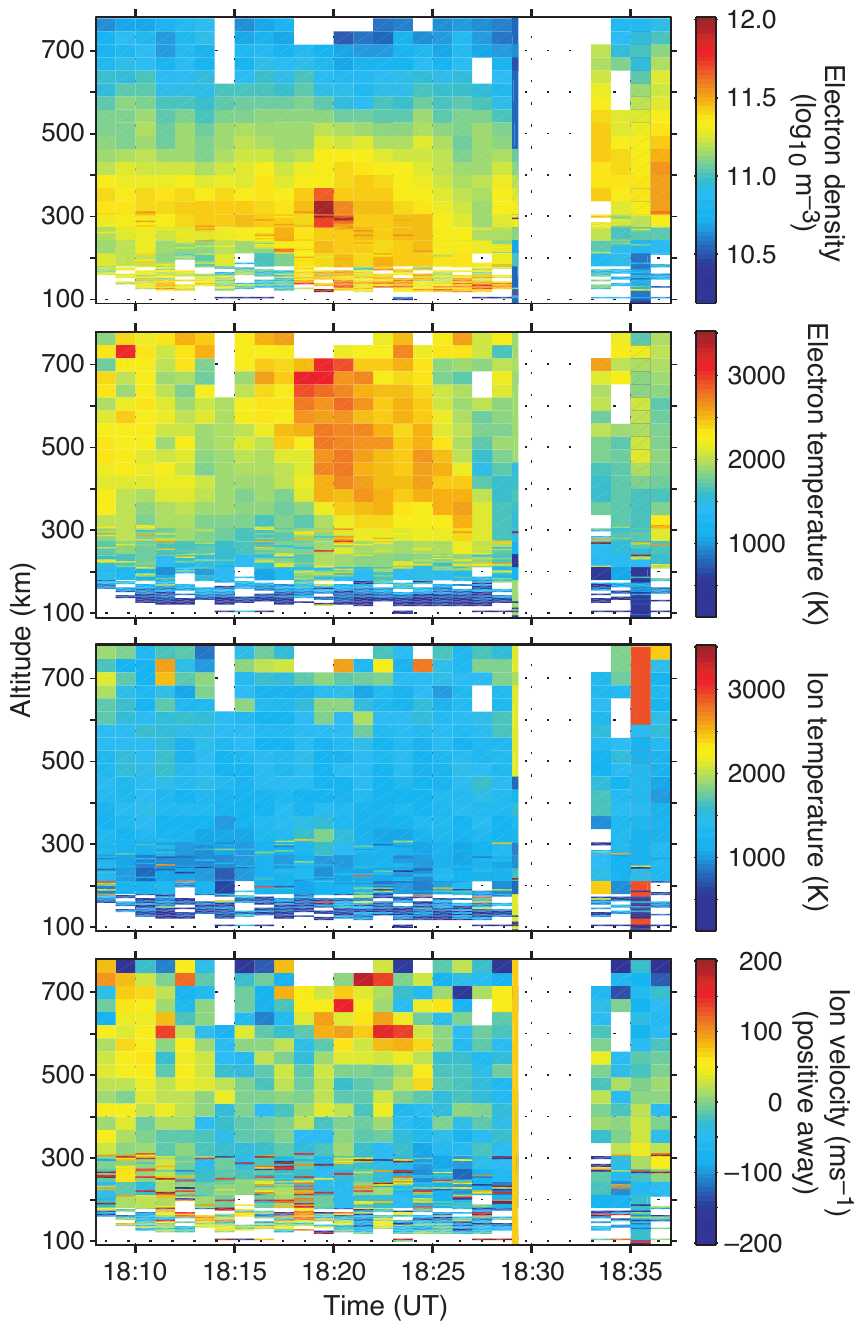}
\caption{ \label{fig:1} (color online)
Background ionospheric parameters measured during the most prominent auroral event.  
The panels show, from top to bottom, 
electron density, 
electron temperature, 
ion temperature, and 
vertical ion velocity (positive indicates motion away from the observer) 
at 1-min time resolution.  
Typical auroral plasma signatures can be seen, namely a sharp increase in electron density localized in time and space, a corresponding increase in electron temperature, localized small increases in ion temperature, and, in the velocity plot, high-altitude ion outflow to space  
\cite{GrydelandA_2004_AnnGeo,BlixtA_2005_AnnGeo_rayed}.  
} \end{figure}

\begin{figure}[t] 
\includegraphics[scale=0.98]{\gpath/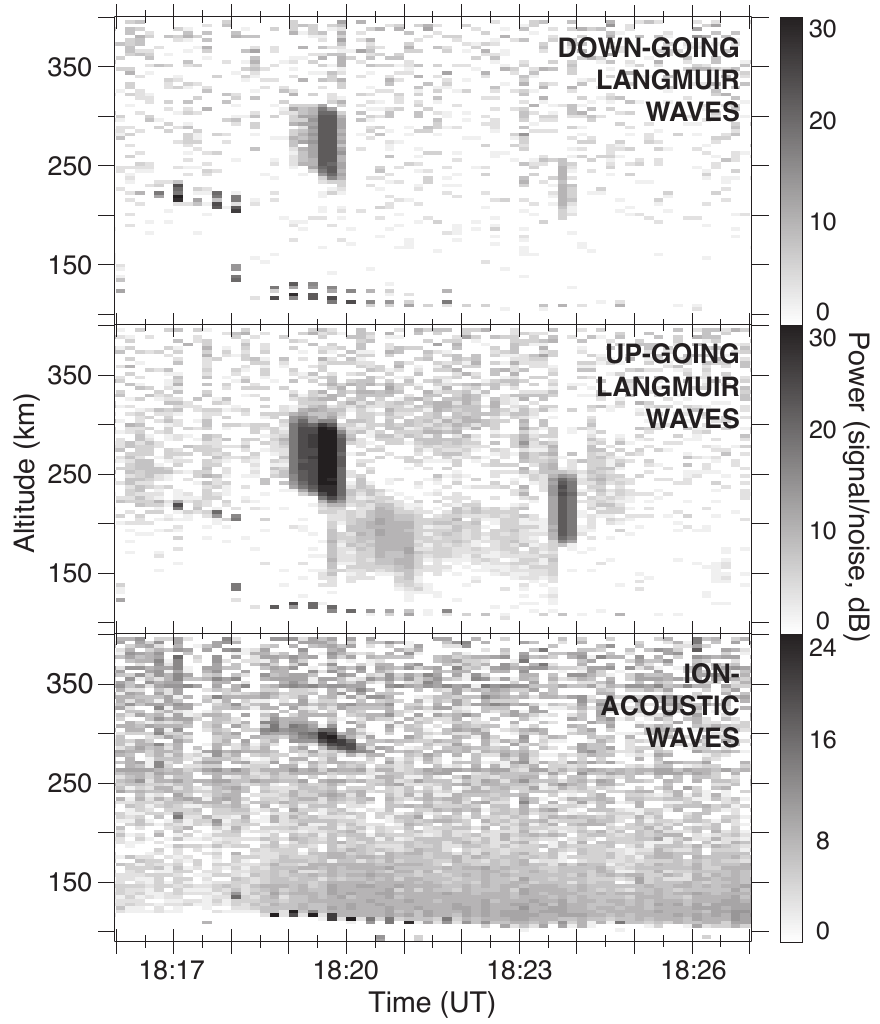}
\caption{ \label{fig:2} 
Incoherent scatter intensity profiles 
from up and down-going Langmuir and ion-acoustic waves 
recorded during the auroral event discussed in Fig.\ 1.  
%
Four distinct sources of scattering can be identified. 
(1) 
The dark background in the ion-acoustic channel is backscatter from thermal-level waves. 
(2) 
The relatively faint bands in the down-going and, more prominently, up-going Langmuir channels 
correspond to 3 and 5 MHz Langmuir waves weakly enhanced by low energy, direct and backscattered diffuse electron precipitation associated with the aurora.  
%
(3) 
The repeated 10-s-long enhancements seen in all three channels at about 225 and 125 km 
before and after 18:18:30, respectively, are backscatter from waves enhanced by 
experimental 
4.04-MHz high-power radiowave transmissions
\cite{RietveldA_2002_ASR}. 
(4) 
The intense features occurring 
between 18:18:30 and 18:20:30 UT near 300 km 
and at 18:23:30 UT near 250 km 
are backscatter associated with the aurora.  
The 
top 
edges of the Langmuir enhancements give the approximate altitudes of the enhanced backscatter. 
%
Uncoded 420-$\mu$s and 25-$\mu$s pulses were used to measure the Langmuir and ion-acoustic backscatter, respectively.
} \end{figure}

\begin{figure}[t] 
\includegraphics[scale=0.9]{\gpath/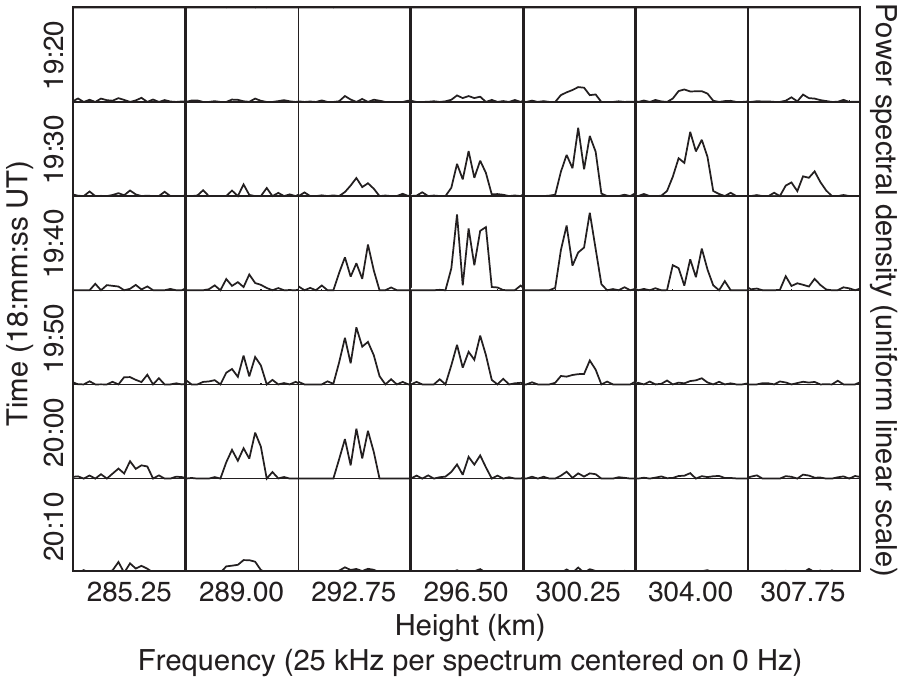}
\caption{ \label{fig:3} 
Power spectral densities of the naturally-enhanced ion-acoustic backscatter 
showing enhanced shoulders and enhanced central peaks 
at 2-kHz frequency and 3.75-km range resolution. 
A 475-$\mu$s phase-coded pulse with a 25-$\mu$s baud was used. 
%
The change in enhancement height  
with time is a result of the vertical pointing of the radar, while the auroral structure is oriented along the geomagnetic field, 
tilted 13${}^\circ$ 
south of vertical, and drifting south; this effect can also be seen in Fig.\ 2.  
The progression from higher to lower heights with time corresponds to a 
drift velocity of about 80 m/s. 
} \end{figure}

Between 18 and 21 UT on 11 and 12 November 1999 
a measurement 
program designed to detect both ion-acoustic and 
Langmuir modes 
was 
run
on the European Incoherent Scatter Scientific Association (EISCAT) 224-MHz radar located near Troms{\o} in northern Norway
(local standard time in Norway is UT plus one hour).
The principal objectives 
were to 
observe 
enhanced waves stimulated by high-power radiowave pumping, 
and, in the event of auroral activity, 
to gather data 
on natural energetic waves 
 \cite{RietveldA_2002_ASR}. 
On both nights conditions were disturbed, and enhanced echoes were detected, 
the strongest being on 11 November between 
18:18:30 and 18:21:30 UT, 
during the passage of an 
aurora through the vertically-directed radar beam.
%
Fig.\ \ref{fig:1} 
presents 
parameters derived from the 
ion-acoustic backscatter 
between 18:15 and 18:28 UT, during the 
most intense auroral event. 
%
Fig.\ \ref{fig:2} 
shows the intensities of Langmuir and ion-acoustic backscatter as a function of height and time. 
The prominent features occurring 
between 18:18:30 and 18:20:30 UT 
and at 18:23:30 UT 
near 300 and 250 km altitude, respectively, 
are backscatter associated with the aurora, 
and are the most energetic natural events observed on either night.  
Two other events occurred later that evening and two more on the following evening.  
Weak ion-acoustic enhancements occurred during each event; 
the Langmuir  
enhancements, however, are always stronger.  
%
Fig.\ \ref{fig:3} 
shows up and down-shifted spectral lines, or ``shoulders'', which are produced by Doppler-shifted backscatter from the 
down and up-going 
ion-acoustic waves, respectively. 
The shoulders are strongly enhanced indicating that the 
waves are nonlinearly amplified. 
In addition, there is a strong central peak, 
a feature not present in thermal-level spectra.

\begin{figure}[t] 
\includegraphics[scale=0.52]{\gpath/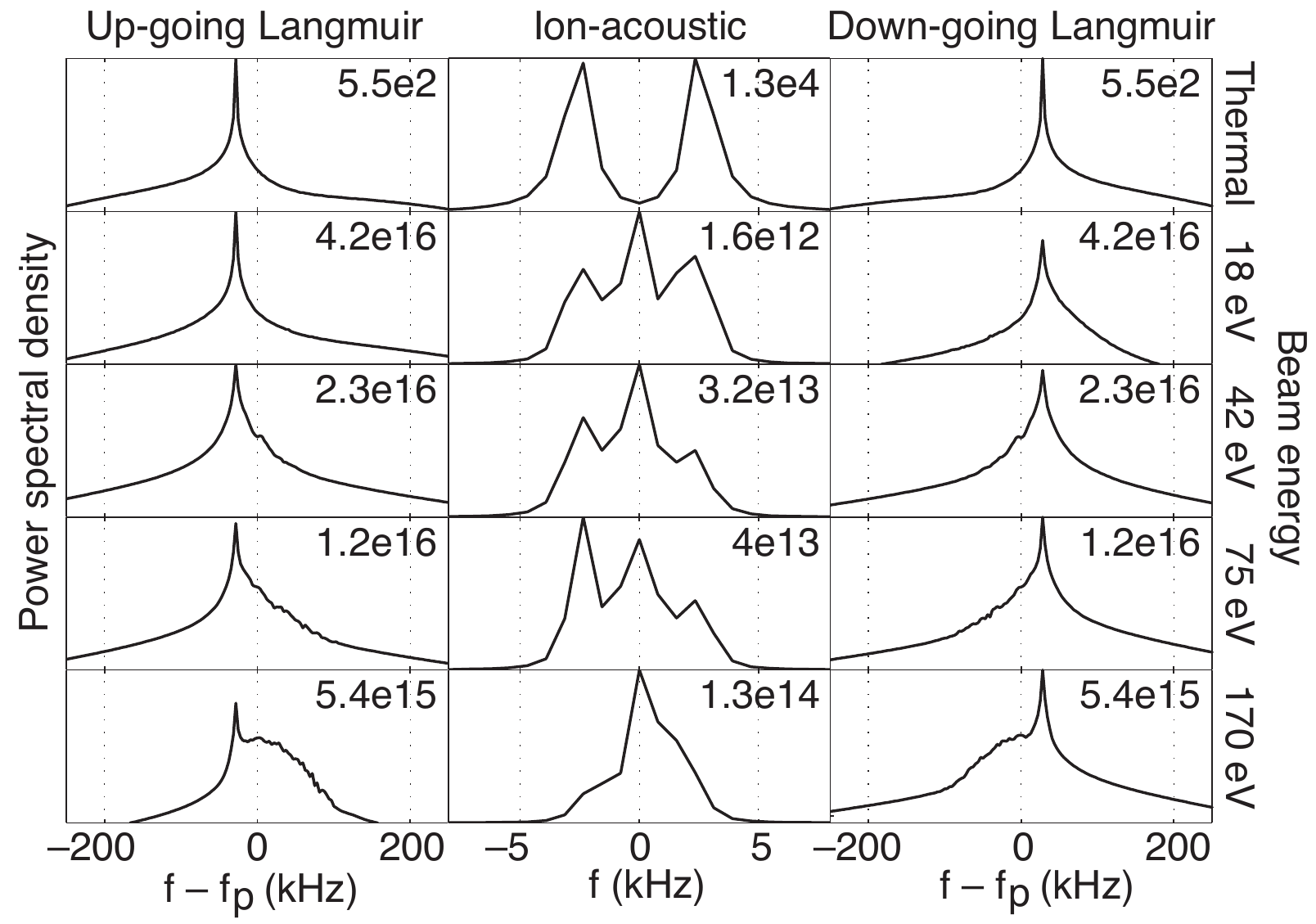}
\caption{ \label{fig:4} 
Results from a 1-d simulation made for plasma parameters matching those 
during this observation and with a downward-going beam of electrons at 
beam energies of 18, 42, 75, and 170 eV. 
The beam creates a bump-on-tail distribution which excites Langmuir waves according to the resonance condition 
$v_{\rm b} \approx {\lambda}_{\rm L} f_{\rm L}$, 
where 
$v_{\rm b}$ is the beam velocity and 
${\lambda}_{\rm L}$ and $f_{\rm L}$ are the Langmuir wavelength and frequency.  
The spectra of 
up and down-going 
Langmuir and ion-acoustic waves were calculated for a wavelength 
of 0.67 m, matching 
the Bragg backscatter condition of the radar.  
Both the precipitating electron beam and the thermal background were included in the driving terms.  
The beam velocity spread ratio ${\Delta}v_{\rm b}/v_{\rm b}$ is 0.3 in all cases, 
where 
${\Delta}v_{\rm b}$ is the velocity spread.  
The beam density ratio $n_{\rm b}/n_{\rm p}$ is 
$2 \cdot 10^{-5}$, 
where $n_{\rm b}$ is the beam density and 
$n_{\rm p}$ is the density of the surrounding plasma, 
$3 \cdot 10^{11}$ m$^{-3}$, 
as determined from the 5-MHz frequency of the 
enhanced Langmuir modes. 
The electron collision frequency is 200 s$^{-1}$.  
These parameters 
are well within the ranges known to occur 
in the auroral ionosphere 
\cite{%
WilliamsA_2001_AnnGeo}.
The peak value of each spectrum is given at the top right of each panel. 
The Langmuir spectra are shown 
on identical log scales 
with 
arbitrary but equal reference levels 
and with 
a minimum value 10$^{-6}$ that of the maximum; 
the ion-acoustic spectra are shown on a linear scale in arbitrary units with a minimum value of zero.
} \end{figure}

The results of a computation 
made for plasma parameters matching those which occurred during this observation are shown in 
Fig.\ \ref{fig:4}. 
A numerical code incorporating a 
one-dimensional periodic version of the Zakharov equations was used 
\cite{%
Zakharov_1972_SovPhysJETP, 
GuioForme_2006_PP}, 
capable of producing the full range of 
cascading (sometimes called weak), 
coexistence (transitional), and 
cavitating (strong) 
Langmuir turbulence.
Energy was supplied by a downward-going flux, or beam, of electrons
\cite{%
WilliamsA_2001_AnnGeo}, 
which excites a Langmuir ``pump'' wave via the bump-on-tail instability. 
In the cascading turbulence scenario 
\cite{%
Forme_1999_AnnGeo}, 
the Langmuir wave then undergoes parametric decay into daughter Langmuir and ion-acoustic waves.  
These waves however exist only within two relatively narrow bands of wave numbers: 
the Langmuir band defined by the driving beam 
(see caption to Fig.\ 4) 
and the ion-acoustic band at about twice that value.  
Furthermore, a radar sees only the wave number that matches the Bragg scattering condition for the radar wavelength. 
This means that, for beam-driven cascading turbulence, 
a radar will see either enhanced Langmuir waves or enhanced ion-acoustic waves, but both may be seen simultaneously only when the velocity spread of the beam is sufficiently broad, approaching the absolute velocity of the beam itself.  
%
Cavitating turbulence is different in that 
enhanced wave modes 
cover a range of $k$ space which extends broadly on both sides of the wave number of the pump wave 
irrespective of the beam velocity breadth.  
In the coexistence or transitional case 
the wave number spectrum extends below the pump wave number to zero, but dies out rapidly for 
Langmuir wave numbers 
greater than the Langmuir pump and for 
ion-acoustic wave numbers 
greater than twice the Langmuir pump.  

In the simulations presented in 
Fig.\ \ref{fig:4}
several different beam energies are modeled, each capable of producing cavitating turbulence, and the spectra are those which would be seen by a radar of the same wavelength as was used for the observations; however, the beam parameters were chosen so that the Langmuir and ion-acoustic wave numbers for cascading turbulence would not match that of the radar.  
%
The coincident enhancement in space and time of both ion-acoustic and Langmuir 
backscatter 
at a single radar wave number 
is a prediction characteristic of cavitating Langmuir turbulence, 
and 
constitutes critical evidence 
for its occurrence 
\cite{
Robinson_1997_RMP, 
GuioForme_2006_PP}.  

A second key feature of this observation 
lies in the shape of the ion-acoustic spectra, which consists of enhanced up and downshifted ``shoulders'' and an enhanced central peak, 
shown in 
Fig.\ \ref{fig:3}. 
The observed spectra match the computed spectra
in
Fig.\ \ref{fig:4}
very well.  
Enhanced shoulders are a standard feature of all past observations of naturally-enhanced ion-acoustic backscatter 
\cite{SedgemoreSchulthessStMaurice_2001_SG}, 
but only one past result shows an enhanced ion-acoustic spectrum with a central peak 
\cite{FormeA_1995_JGR}.  
The appearance of the non-Doppler-shifted central peak indicates the presence of meter-scale, non-propagating density wells known as cavitons.
A central peak 
is not a necessary 
feature of 
cavitating turbulence 
but will occur when the spacing between the cavitons matches the Bragg condition of the radar.  
The spacing, in turn, is roughly proportional to the inverse square root of the energy density 
of the primary Langmuir waves, 
or, in turn,
of the pump beam 
\citep{
ShenNicholson_1987_PhysFluids, 
GuioForme_2006_PP}.  
This is not likely to be seen in all observations, both because of the matching requirement and because a relatively high beam energy density is required, which is most likely to be observable using long wavelengths such as in the observations reported here and 
previously 
  \cite{FormeA_1995_JGR}. 

Many other features of the observations may also be explained by the cavitating 
turbulence model.  
%
%
(1) Six events were observed, four on the 11th and two on the 12th, and both Langmuir and ion-acoustic enhancements exist in all cases, but in all events the ion-acoustic enhancement is weaker.  
This agrees with 
the computed examples. 
%
(2) The measured 
backscatter intensity in the down-going Langmuir channels is 
weaker 
than 
the up-going. 
Similar differences can be seen in the computed spectra. 
For the case of 18 eV, 
the radar 
sees 
down-going 
Langmuir modes 
at a wave number 
corresponding 
to the negative, or damping, slope of the velocity distribution function of the down-going beam, and heavy damping is seen.  
For wave numbers 
on the positive slope of the beam 
the opposite will occur.  
%
%
(3) In five of our cases the ion-acoustic enhancement disappears before, or at the same time as, 
the Langmuir enhancements 
(within the 10-s time resolution of the observations).  
However, in our strongest event, at 18:18:30 on the 11th, enhanced ion-acoustic backscatter is seen before and after the enhanced Langmuir backscatter.  
In this case it is possible that the turbulence 
develops 
from coexistence to cavitating and back again as the driving beam grows and then decays.  
In the coexistence regime 
Langmuir mode wavenumbers 
are cut off at a value roughly half that 
of the ion-acoustic mode, 
so 
the radar 
may see only the ion-acoustic enhancement 
until the cavitating turbulence is fully developed.  
A strong event would also cause greater electron heating, 
reducing damping of ion-acoustic modes 
and contributing to a longer ion-acoustic enhancement.  
(4) 
The relationship between beam energy and ion-acoustic damping can also account for the relatively strong ion-acoustic enhancement after 
18:20:00~UT: 
the electrons had 
been heated 
during the event 
and the ion-acoustic damping rate reduced, 
allowing the ion-acoustic backscatter 
to remain strong even as the drive 
began 
to weaken.  

A significant 
feature in the computed Langmuir spectrum 
at 170 eV, and very weakly at 42 and 75 eV, is 
the appearance of a broad spectrum 
at and near 
the cold plasma frequency.  
This feature, which is not resolvable in the radar measurements presented here, 
is due to Langmuir waves trapped in cavitons. 
It appears after a sufficient period of pumping at a sufficiently high level. 
The central peak and the broad spectrum, both due to cavitation, 
may be seen 
under somewhat different circumstances: 
The central peak requires 
a beam energy density 
that allows 
the caviton spacing to match the radar Bragg condition, 
while 
the broad spectrum requires 
a fixed electron density with 
a beam velocity high enough 
and beam duration long enough 
to allow formation of trapped Langmuir waves matching the radar Bragg condition.  
Both are seen in high-power radiowave experiments 
 \cite{
 RietveldA_2000_JGR}, 
in which a fixed pump frequency substitutes for a fixed electron density. 

The data presented here provide the first direct evidence of naturally occurring cavitating Langmuir turbulence, 
thought to be important in 
space and astrophysical plasmas as varied as pulsar magnetospheres and the Earth's ionosphere 
\cite{Robinson_1997_RMP, 
AsseoPorzi_2006_MNRAS,
NulsenA_2007_JGR,
KelloggA_1992_GRL, 
RobinsonCairns_1995_GRL, 
StasiewiczA_1996_JGR, 
PapadopoulosCoffey_1974_JGR_nonthermal, 
RowlandA_1981_PRL, 
GuioForme_2006_PP}. 
Further observations 
of 
Langmuir turbulence 
in the ionosphere may yield advances in our understanding of
supra-thermal electron distributions \cite{MatthewsA_1976_JGR}, 
naturally-enhanced ion-acoustic waves \cite{Forme_1999_AnnGeo,GuioForme_2006_PP}, 
natural ionospheric radio emissions \cite{HughesLaBelle_2001_JGR},
anomalous resistivity \cite{PapadopoulosCoffey_1974_JGR_anomalous,RowlandA_1981_GRL}, 
and 
auroral currents and dark aurora \cite{GrydelandA_2004_AnnGeo}.

\begin{acknowledgments}
EISCAT 
is supported by research organizations in 
China, Finland, France, Germany, Japan, Norway, Russia, Sweden, Ukraine, and the 
UK. %
%
Support for B.I.\ and M.T.R.\ was provided in part, respectively, 
by U.S.\ Army Research Office contract W911NF-07-1-0016
and 
by the Max-Planck-Institute for Aero\-no\-my. 
Some of the calculations were performed on facilities provided by the Miracle Consortium, part of the DiRAC project funded by the UK Science and Technology Facilities Council.
We thank 
C{\'e}sar La Hoz for helping to arrange observation time on the EISCAT radar system, 
and we thank 
C{\'e}sar La Hoz, 
Jim LaBelle, 
Kristina Lynch, 
Iver Cairns, 
Gerhard Haerendal, 
and two anonymous referees for constructive and helpful comments. 
\end{acknowledgments}


\begin{thebibliography}{29}%
\makeatletter
\providecommand \@ifxundefined [1]{%
 \@ifx{#1\undefined}
}%
\providecommand \@ifnum [1]{%
 \ifnum #1\expandafter \@firstoftwo
 \else \expandafter \@secondoftwo
 \fi
}%
\providecommand \@ifx [1]{%
 \ifx #1\expandafter \@firstoftwo
 \else \expandafter \@secondoftwo
 \fi
}%
\providecommand \natexlab [1]{#1}%
\providecommand \enquote  [1]{``#1''}%
\providecommand \bibnamefont  [1]{#1}%
\providecommand \bibfnamefont [1]{#1}%
\providecommand \citenamefont [1]{#1}%
\providecommand \href@noop [0]{\@secondoftwo}%
\providecommand \href [0]{\begingroup \@sanitize@url \@href}%
\providecommand \@href[1]{\@@startlink{#1}\@@href}%
\providecommand \@@href[1]{\endgroup#1\@@endlink}%
\providecommand \@sanitize@url [0]{\catcode `\\12\catcode `\$12\catcode
  `\&12\catcode `\#12\catcode `\^12\catcode `\_12\catcode `\%12\relax}%
\providecommand \@@startlink[1]{}%
\providecommand \@@endlink[0]{}%
\providecommand \url  [0]{\begingroup\@sanitize@url \@url }%
\providecommand \@url [1]{\endgroup\@href {#1}{\urlprefix }}%
\providecommand \urlprefix  [0]{URL }%
\providecommand \Eprint [0]{\href }%
\@ifxundefined \urlstyle {%
  \providecommand \doi  [0]{\begingroup \@sanitize@url \@doi}%
  \providecommand \@doi [1]{\endgroup \@@startlink {\doibase
  #1}doi:\discretionary {}{}{}#1\@@endlink }%
}{%
  \providecommand \doi  [0]{doi:\discretionary{}{}{}\begingroup
  \urlstyle{rm}\Url }%
}%
\providecommand \doibase [0]{http://dx.doi.org/}%
\providecommand \Doi [0]{\begingroup \@sanitize@url \@Doi }%
\providecommand \@Doi  [1]{\endgroup\@@startlink{\doibase#1}\@@Doi}%
\providecommand \@@Doi [1]{#1\@@endlink}%
\providecommand \selectlanguage [0]{\@gobble}%
\providecommand \bibinfo  [0]{\@secondoftwo}%
\providecommand \bibfield  [0]{\@secondoftwo}%
\providecommand \translation [1]{[#1]}%
\providecommand \BibitemOpen [0]{}%
\providecommand \bibitemStop [0]{}%
\providecommand \bibitemNoStop [0]{.\EOS\space}%
\providecommand \EOS [0]{\spacefactor3000\relax}%
\providecommand \BibitemShut  [1]{\csname bibitem#1\endcsname}%
\bibitem [{\citenamefont {Wong}\ and\ \citenamefont
  {Cheung}(1984)}]{WongCheung_1984_PRL}%
  \BibitemOpen
  \bibfield  {author} {\bibinfo {author} {\bibfnamefont {A.~Y.}\ \bibnamefont
  {Wong}}\ and\ \bibinfo {author} {\bibfnamefont {P.~Y.}\ \bibnamefont
  {Cheung}},\ }\Doi {10.1103/PhysRevLett.52.1222} {\bibfield  {journal}
  {\bibinfo  {journal} {Phys. Rev. Lett.},\ }\textbf {\bibinfo {volume} {52}},\
  \bibinfo {pages} {1222} (\bibinfo {year} {1984})}\BibitemShut {NoStop}%
\bibitem [{\citenamefont {{Vyacheslavov}}\ \emph {et~al.}(2002)\citenamefont
  {{Vyacheslavov}}, \citenamefont {{Burmasov}}, \citenamefont {{Kandaurov}},
  \citenamefont {{Kruglyakov}}, \citenamefont {{Meshkov}}, \citenamefont
  {{Popov}},\ and\ \citenamefont {{Sanin}}}]{VyacheslavovA_2002_PlPhCF}%
  \BibitemOpen
  \bibfield  {author} {\bibinfo {author} {\bibfnamefont {L.~N.}\ \bibnamefont
  {{Vyacheslavov}}}, \bibinfo {author} {\bibfnamefont {V.~S.}\ \bibnamefont
  {{Burmasov}}}, \bibinfo {author} {\bibfnamefont {I.~V.}\ \bibnamefont
  {{Kandaurov}}}, \bibinfo {author} {\bibfnamefont {E.~P.}\ \bibnamefont
  {{Kruglyakov}}}, \bibinfo {author} {\bibfnamefont {O.~I.}\ \bibnamefont
  {{Meshkov}}}, \bibinfo {author} {\bibfnamefont {S.~S.}\ \bibnamefont
  {{Popov}}}, \ and\ \bibinfo {author} {\bibfnamefont {A.~L.}\ \bibnamefont
  {{Sanin}}},\ }\href@noop {} {\bibfield  {journal} {\bibinfo  {journal}
  {Plasma Phys. Control. Fusion},\ }\textbf {\bibinfo {volume} {44}},\ \bibinfo
  {pages} {B279+} (\bibinfo {year} {2002})}\BibitemShut {NoStop}%
\bibitem [{\citenamefont {Sulzer}\ and\ \citenamefont
  {Fejer}(1994)}]{SulzerFejer_1994_JGR}%
  \BibitemOpen
  \bibfield  {author} {\bibinfo {author} {\bibfnamefont {M.~P.}\ \bibnamefont
  {Sulzer}}\ and\ \bibinfo {author} {\bibfnamefont {J.~A.}\ \bibnamefont
  {Fejer}},\ }\href@noop {} {\bibfield  {journal} {\bibinfo  {journal} {J.
  Geophys. Res.},\ }\textbf {\bibinfo {volume} {99}},\ \bibinfo {pages} {15035}
  (\bibinfo {year} {1994})}\BibitemShut {NoStop}%
\bibitem [{\citenamefont {Isham}\ \emph {et~al.}(1999)\citenamefont {Isham},
  \citenamefont {{La Hoz}}, \citenamefont {Rietveld}, \citenamefont {Hagfors},\
  and\ \citenamefont {Leyser}}]{IshamA_1999_PRL}%
  \BibitemOpen
  \bibfield  {author} {\bibinfo {author} {\bibfnamefont {B.}~\bibnamefont
  {Isham}}, \bibinfo {author} {\bibfnamefont {C.}~\bibnamefont {{La Hoz}}},
  \bibinfo {author} {\bibfnamefont {M.~T.}\ \bibnamefont {Rietveld}}, \bibinfo
  {author} {\bibfnamefont {T.}~\bibnamefont {Hagfors}}, \ and\ \bibinfo
  {author} {\bibfnamefont {T.~B.}\ \bibnamefont {Leyser}},\ }\Doi
  {10.1103/PhysRevLett.83.2576} {\bibfield  {journal} {\bibinfo  {journal}
  {Phys. Rev. Lett.},\ }\textbf {\bibinfo {volume} {83}},\ \bibinfo {pages}
  {2576} (\bibinfo {year} {1999})}\BibitemShut {NoStop}%
\bibitem [{\citenamefont {Rietveld}\ \emph {et~al.}(2000)\citenamefont
  {Rietveld}, \citenamefont {Isham}, \citenamefont {Kohl}, \citenamefont {{La
  Hoz}},\ and\ \citenamefont {Hagfors}}]{RietveldA_2000_JGR}%
  \BibitemOpen
  \bibfield  {author} {\bibinfo {author} {\bibfnamefont {M.~T.}\ \bibnamefont
  {Rietveld}}, \bibinfo {author} {\bibfnamefont {B.}~\bibnamefont {Isham}},
  \bibinfo {author} {\bibfnamefont {H.}~\bibnamefont {Kohl}}, \bibinfo {author}
  {\bibfnamefont {C.}~\bibnamefont {{La Hoz}}}, \ and\ \bibinfo {author}
  {\bibfnamefont {T.}~\bibnamefont {Hagfors}},\ }\href@noop {} {\bibfield
  {journal} {\bibinfo  {journal} {J. Geophys. Res.},\ }\textbf {\bibinfo
  {volume} {105}},\ \bibinfo {pages} {7429} (\bibinfo {year}
  {2000})}\BibitemShut {NoStop}%
\bibitem [{\citenamefont {{Asseo}}\ and\ \citenamefont
  {{Porzio}}(2006)}]{AsseoPorzi_2006_MNRAS}%
  \BibitemOpen
  \bibfield  {author} {\bibinfo {author} {\bibfnamefont {E.}~\bibnamefont
  {{Asseo}}}\ and\ \bibinfo {author} {\bibfnamefont {A.}~\bibnamefont
  {{Porzio}}},\ }\Doi {10.1111/j.1365-2966.2006.10386.x} {\bibfield  {journal}
  {\bibinfo  {journal} {Mon. Not. R. Astr.Soc},\ }\textbf {\bibinfo {volume}
  {369}},\ \bibinfo {pages} {1469} (\bibinfo {year} {2006})}\BibitemShut
  {NoStop}%
\bibitem [{\citenamefont {{Nulsen}}\ \emph {et~al.}(2007)\citenamefont
  {{Nulsen}}, \citenamefont {{Cairns}},\ and\ \citenamefont
  {{Robinson}}}]{NulsenA_2007_JGR}%
  \BibitemOpen
  \bibfield  {author} {\bibinfo {author} {\bibfnamefont {A.~L.}\ \bibnamefont
  {{Nulsen}}}, \bibinfo {author} {\bibfnamefont {I.~H.}\ \bibnamefont
  {{Cairns}}}, \ and\ \bibinfo {author} {\bibfnamefont {P.~A.}\ \bibnamefont
  {{Robinson}}},\ }\Doi {10.1029/2006JA011873} {\bibfield  {journal} {\bibinfo
  {journal} {J. Geophys. Res.},\ }\textbf {\bibinfo {volume} {112}},\ \bibinfo
  {pages} {5107} (\bibinfo {year} {2007})}\BibitemShut {NoStop}%
\bibitem [{\citenamefont {{Kellogg}}\ \emph {et~al.}(1992)\citenamefont
  {{Kellogg}}, \citenamefont {{Goetz}}, \citenamefont {{Howard}},\ and\
  \citenamefont {{Monson}}}]{KelloggA_1992_GRL}%
  \BibitemOpen
  \bibfield  {author} {\bibinfo {author} {\bibfnamefont {P.~J.}\ \bibnamefont
  {{Kellogg}}}, \bibinfo {author} {\bibfnamefont {K.}~\bibnamefont {{Goetz}}},
  \bibinfo {author} {\bibfnamefont {R.~L.}\ \bibnamefont {{Howard}}}, \ and\
  \bibinfo {author} {\bibfnamefont {S.~J.}\ \bibnamefont {{Monson}}},\
  }\href@noop {} {\bibfield  {journal} {\bibinfo  {journal} {Geophys. Res.
  Lett.},\ }\textbf {\bibinfo {volume} {19}},\ \bibinfo {pages} {1303}
  (\bibinfo {year} {1992})}\BibitemShut {NoStop}%
\bibitem [{\citenamefont {{Robinson}}\ and\ \citenamefont
  {{Cairns}}(1995)}]{RobinsonCairns_1995_GRL}%
  \BibitemOpen
  \bibfield  {author} {\bibinfo {author} {\bibfnamefont {P.~A.}\ \bibnamefont
  {{Robinson}}}\ and\ \bibinfo {author} {\bibfnamefont {I.~H.}\ \bibnamefont
  {{Cairns}}},\ }\Doi {10.1029/95GL01779} {\bibfield  {journal} {\bibinfo
  {journal} {Geophys. Res. Lett.},\ }\textbf {\bibinfo {volume} {22}},\
  \bibinfo {pages} {2657} (\bibinfo {year} {1995})}\BibitemShut {NoStop}%
\bibitem [{\citenamefont {{Stasiewicz}}\ \emph {et~al.}(1996)\citenamefont
  {{Stasiewicz}}, \citenamefont {{Holback}}, \citenamefont {{Krasnoselskikh}},
  \citenamefont {{Boehm}}, \citenamefont {{Bostr{\"o}m}},\ and\ \citenamefont
  {{Kintner}}}]{StasiewiczA_1996_JGR}%
  \BibitemOpen
  \bibfield  {author} {\bibinfo {author} {\bibfnamefont {K.}~\bibnamefont
  {{Stasiewicz}}}, \bibinfo {author} {\bibfnamefont {B.}~\bibnamefont
  {{Holback}}}, \bibinfo {author} {\bibfnamefont {V.}~\bibnamefont
  {{Krasnoselskikh}}}, \bibinfo {author} {\bibfnamefont {M.}~\bibnamefont
  {{Boehm}}}, \bibinfo {author} {\bibfnamefont {R.}~\bibnamefont
  {{Bostr{\"o}m}}}, \ and\ \bibinfo {author} {\bibfnamefont {P.~M.}\
  \bibnamefont {{Kintner}}},\ }\href@noop {} {\bibfield  {journal} {\bibinfo
  {journal} {J. Geophys. Res.},\ }\textbf {\bibinfo {volume} {101}},\ \bibinfo
  {pages} {21515} (\bibinfo {year} {1996})}\BibitemShut {NoStop}%
\bibitem [{\citenamefont {{Papadopoulos}}\ and\ \citenamefont
  {{Coffey}}(1974){\natexlab{a}}}]{PapadopoulosCoffey_1974_JGR_nonthermal}%
  \BibitemOpen
  \bibfield  {author} {\bibinfo {author} {\bibfnamefont {K.}~\bibnamefont
  {{Papadopoulos}}}\ and\ \bibinfo {author} {\bibfnamefont {T.}~\bibnamefont
  {{Coffey}}},\ }\Doi {10.1029/JA079i004p00674} {\bibfield  {journal} {\bibinfo
   {journal} {J. Geophys. Res.},\ }\textbf {\bibinfo {volume} {79}},\ \bibinfo
  {pages} {674} (\bibinfo {year} {1974}{\natexlab{a}})}\BibitemShut {NoStop}%
\bibitem [{\citenamefont {{Rowland}}\ \emph
  {et~al.}(1981){\natexlab{a}}\citenamefont {{Rowland}}, \citenamefont
  {{Lyon}},\ and\ \citenamefont {{Papadopoulos}}}]{RowlandA_1981_PRL}%
  \BibitemOpen
  \bibfield  {author} {\bibinfo {author} {\bibfnamefont {H.~L.}\ \bibnamefont
  {{Rowland}}}, \bibinfo {author} {\bibfnamefont {J.~G.}\ \bibnamefont
  {{Lyon}}}, \ and\ \bibinfo {author} {\bibfnamefont {K.}~\bibnamefont
  {{Papadopoulos}}},\ }\Doi {10.1103/PhysRevLett.46.346} {\bibfield  {journal}
  {\bibinfo  {journal} {Phys. Rev. Lett.},\ }\textbf {\bibinfo {volume} {46}},\
  \bibinfo {pages} {346} (\bibinfo {year} {1981}{\natexlab{a}})}\BibitemShut
  {NoStop}%
\bibitem [{\citenamefont {{Guio}}\ and\ \citenamefont
  {{Forme}}(2006)}]{GuioForme_2006_PP}%
  \BibitemOpen
  \bibfield  {author} {\bibinfo {author} {\bibfnamefont {P.}~\bibnamefont
  {{Guio}}}\ and\ \bibinfo {author} {\bibfnamefont {F.}~\bibnamefont
  {{Forme}}},\ }\Doi {10.1063/1.2402145} {\bibfield  {journal} {\bibinfo
  {journal} {Phys. Plasmas},\ }\textbf {\bibinfo {volume} {13}},\ \bibinfo
  {pages} {(122902} (\bibinfo {year} {2006})}\BibitemShut {NoStop}%
\bibitem [{\citenamefont {Robinson}(1997)}]{Robinson_1997_RMP}%
  \BibitemOpen
  \bibfield  {author} {\bibinfo {author} {\bibfnamefont {P.~A.}\ \bibnamefont
  {Robinson}},\ }\Doi {10.1103/RevModPhys.69.507} {\bibfield  {journal}
  {\bibinfo  {journal} {Rev. Mod. Phys.},\ }\textbf {\bibinfo {volume} {69}},\
  \bibinfo {pages} {507} (\bibinfo {year} {1997})}\BibitemShut {NoStop}%
\bibitem [{\citenamefont {{DuBois}}\ \emph {et~al.}(1991)\citenamefont
  {{DuBois}}, \citenamefont {{Rose}},\ and\ \citenamefont
  {{Russell}}}]{DuBoisA_1991_PRL}%
  \BibitemOpen
  \bibfield  {author} {\bibinfo {author} {\bibfnamefont {D.~F.}\ \bibnamefont
  {{DuBois}}}, \bibinfo {author} {\bibfnamefont {H.~A.}\ \bibnamefont
  {{Rose}}}, \ and\ \bibinfo {author} {\bibfnamefont {D.}~\bibnamefont
  {{Russell}}},\ }\href@noop {} {\bibfield  {journal} {\bibinfo  {journal}
  {Phys. Rev. Lett.},\ }\textbf {\bibinfo {volume} {66}},\ \bibinfo {pages}
  {1970} (\bibinfo {year} {1991})}\BibitemShut {NoStop}%
\bibitem [{\citenamefont {{DuBois}}\ \emph {et~al.}(1993)\citenamefont
  {{DuBois}}, \citenamefont {Hanssen}, \citenamefont {Rose},\ and\
  \citenamefont {Russell}}]{DuBoisA_1993_PhysFluids}%
  \BibitemOpen
  \bibfield  {author} {\bibinfo {author} {\bibfnamefont {D.~F.}\ \bibnamefont
  {{DuBois}}}, \bibinfo {author} {\bibfnamefont {A.}~\bibnamefont {Hanssen}},
  \bibinfo {author} {\bibfnamefont {H.~A.}\ \bibnamefont {Rose}}, \ and\
  \bibinfo {author} {\bibfnamefont {D.}~\bibnamefont {Russell}},\ }\href@noop
  {} {\bibfield  {journal} {\bibinfo  {journal} {Phys. Fluids},\ }\textbf
  {\bibinfo {volume} {B5}},\ \bibinfo {pages} {2616} (\bibinfo {year}
  {1993})}\BibitemShut {NoStop}%
\bibitem [{\citenamefont {Rietveld}\ \emph {et~al.}(2002)\citenamefont
  {Rietveld}, \citenamefont {Isham}, \citenamefont {Grydeland}, \citenamefont
  {{La Hoz}}, \citenamefont {Leyser}, \citenamefont {Honary}, \citenamefont
  {Ueda}, \citenamefont {Kosch},\ and\ \citenamefont
  {Hagfors}}]{RietveldA_2002_ASR}%
  \BibitemOpen
  \bibfield  {author} {\bibinfo {author} {\bibfnamefont {M.~T.}\ \bibnamefont
  {Rietveld}}, \bibinfo {author} {\bibfnamefont {B.}~\bibnamefont {Isham}},
  \bibinfo {author} {\bibfnamefont {T.}~\bibnamefont {Grydeland}}, \bibinfo
  {author} {\bibfnamefont {C.}~\bibnamefont {{La Hoz}}}, \bibinfo {author}
  {\bibfnamefont {T.~B.}\ \bibnamefont {Leyser}}, \bibinfo {author}
  {\bibfnamefont {F.}~\bibnamefont {Honary}}, \bibinfo {author} {\bibfnamefont
  {H.}~\bibnamefont {Ueda}}, \bibinfo {author} {\bibfnamefont {M.}~\bibnamefont
  {Kosch}}, \ and\ \bibinfo {author} {\bibfnamefont {T.}~\bibnamefont
  {Hagfors}},\ }\Doi {10.1016/S0273-1177(02)00186-2} {\bibfield  {journal}
  {\bibinfo  {journal} {Adv. Space Res.},\ }\textbf {\bibinfo {volume} {29}},\
  \bibinfo {pages} {1363} (\bibinfo {year} {2002})}\BibitemShut {NoStop}%
\bibitem [{\citenamefont {Grydeland}\ \emph {et~al.}(2004)\citenamefont
  {Grydeland}, \citenamefont {Blixt}, \citenamefont {L{\o}vhaug}, \citenamefont
  {Hagfors}, \citenamefont {Hoz},\ and\ \citenamefont
  {Trondsen}}]{GrydelandA_2004_AnnGeo}%
  \BibitemOpen
  \bibfield  {author} {\bibinfo {author} {\bibfnamefont {T.}~\bibnamefont
  {Grydeland}}, \bibinfo {author} {\bibfnamefont {E.~M.}\ \bibnamefont
  {Blixt}}, \bibinfo {author} {\bibfnamefont {U.~P.}\ \bibnamefont
  {L{\o}vhaug}}, \bibinfo {author} {\bibfnamefont {T.}~\bibnamefont {Hagfors}},
  \bibinfo {author} {\bibfnamefont {C.~L.}\ \bibnamefont {Hoz}}, \ and\
  \bibinfo {author} {\bibfnamefont {T.~S.}\ \bibnamefont {Trondsen}},\
  }\href@noop {} {\bibfield  {journal} {\bibinfo  {journal} {Ann.\
  Geophysicae},\ }\textbf {\bibinfo {volume} {22}},\ \bibinfo {pages} {1115}
  (\bibinfo {year} {2004})}\BibitemShut {NoStop}%
\bibitem [{\citenamefont {Blixt}\ \emph {et~al.}(2005)\citenamefont {Blixt},
  \citenamefont {Grydeland}, \citenamefont {Ivchenko}, \citenamefont {Hagfors},
  \citenamefont {Hoz}, \citenamefont {Lanchester}, \citenamefont {L{\o}vhaug},\
  and\ \citenamefont {Trondsen}}]{BlixtA_2005_AnnGeo_rayed}%
  \BibitemOpen
  \bibfield  {author} {\bibinfo {author} {\bibfnamefont {E.~M.}\ \bibnamefont
  {Blixt}}, \bibinfo {author} {\bibfnamefont {T.}~\bibnamefont {Grydeland}},
  \bibinfo {author} {\bibfnamefont {N.}~\bibnamefont {Ivchenko}}, \bibinfo
  {author} {\bibfnamefont {T.}~\bibnamefont {Hagfors}}, \bibinfo {author}
  {\bibfnamefont {C.~L.}\ \bibnamefont {Hoz}}, \bibinfo {author} {\bibfnamefont
  {B.~S.}\ \bibnamefont {Lanchester}}, \bibinfo {author} {\bibfnamefont
  {U.~P.}\ \bibnamefont {L{\o}vhaug}}, \ and\ \bibinfo {author} {\bibfnamefont
  {T.~S.}\ \bibnamefont {Trondsen}},\ }\href@noop {} {\bibfield  {journal}
  {\bibinfo  {journal} {Ann.\ Geophysicae},\ }\textbf {\bibinfo {volume}
  {23}},\ \bibinfo {pages} {3} (\bibinfo {year} {2005})}\BibitemShut {NoStop}%
\bibitem [{\citenamefont {Zakharov}(1972)}]{Zakharov_1972_SovPhysJETP}%
  \BibitemOpen
  \bibfield  {author} {\bibinfo {author} {\bibfnamefont {V.~E.}\ \bibnamefont
  {Zakharov}},\ }\href@noop {} {\bibfield  {journal} {\bibinfo  {journal} {Sov.
  Phys. JETP {(English transl.)}},\ }\textbf {\bibinfo {volume} {35}},\
  \bibinfo {pages} {908} (\bibinfo {year} {1972})}\BibitemShut {NoStop}%
\bibitem [{\citenamefont {{Williams}}\ \emph {et~al.}(2006)\citenamefont
  {{Williams}}, \citenamefont {{MacDonald}}, \citenamefont {{McCarthy}},
  \citenamefont {Peticolas},\ and\ \citenamefont
  {{Parks}}}]{WilliamsA_2001_AnnGeo}%
  \BibitemOpen
  \bibfield  {author} {\bibinfo {author} {\bibfnamefont {J.~D.}\ \bibnamefont
  {{Williams}}}, \bibinfo {author} {\bibfnamefont {E.}~\bibnamefont
  {{MacDonald}}}, \bibinfo {author} {\bibfnamefont {M.}~\bibnamefont
  {{McCarthy}}}, \bibinfo {author} {\bibfnamefont {L.}~\bibnamefont
  {Peticolas}}, \ and\ \bibinfo {author} {\bibfnamefont {G.~K.}\ \bibnamefont
  {{Parks}}},\ }\href@noop {} {\bibfield  {journal} {\bibinfo  {journal} {Ann.\
  Geophysicae},\ }\textbf {\bibinfo {volume} {24}},\ \bibinfo {pages} {1829}
  (\bibinfo {year} {2006})}\BibitemShut {NoStop}%
\bibitem [{\citenamefont {{Forme}}(1999)}]{Forme_1999_AnnGeo}%
  \BibitemOpen
  \bibfield  {author} {\bibinfo {author} {\bibfnamefont {F.~R.~E.}\
  \bibnamefont {{Forme}}},\ }\href@noop {} {\bibfield  {journal} {\bibinfo
  {journal} {Ann.\ Geophysicae},\ }\textbf {\bibinfo {volume} {17}},\ \bibinfo
  {pages} {1172} (\bibinfo {year} {1999})}\BibitemShut {NoStop}%
\bibitem [{\citenamefont {Sedgemore-Schulthess}\ and\ \citenamefont
  {St.-Maurice}(2001)}]{SedgemoreSchulthessStMaurice_2001_SG}%
  \BibitemOpen
  \bibfield  {author} {\bibinfo {author} {\bibfnamefont {F.}~\bibnamefont
  {Sedgemore-Schulthess}}\ and\ \bibinfo {author} {\bibfnamefont {J.-P.}\
  \bibnamefont {St.-Maurice}},\ }\href@noop {} {\bibfield  {journal} {\bibinfo
  {journal} {Sur. Geophys.},\ }\textbf {\bibinfo {volume} {22}},\ \bibinfo
  {pages} {55} (\bibinfo {year} {2001})}\BibitemShut {NoStop}%
\bibitem [{\citenamefont {{Forme}}\ \emph {et~al.}(1995)\citenamefont
  {{Forme}}, \citenamefont {{Fontaine}},\ and\ \citenamefont
  {{Wahlund}}}]{FormeA_1995_JGR}%
  \BibitemOpen
  \bibfield  {author} {\bibinfo {author} {\bibfnamefont {F.~R.~E.}\
  \bibnamefont {{Forme}}}, \bibinfo {author} {\bibfnamefont {D.}~\bibnamefont
  {{Fontaine}}}, \ and\ \bibinfo {author} {\bibfnamefont {J.~E.}\ \bibnamefont
  {{Wahlund}}},\ }\href@noop {} {\bibfield  {journal} {\bibinfo  {journal} {J.
  Geophys. Res.},\ }\textbf {\bibinfo {volume} {100}},\ \bibinfo {pages}
  {14625} (\bibinfo {year} {1995})}\BibitemShut {NoStop}%
\bibitem [{\citenamefont {Shen}\ and\ \citenamefont
  {Nicholson}(1987)}]{ShenNicholson_1987_PhysFluids}%
  \BibitemOpen
  \bibfield  {author} {\bibinfo {author} {\bibfnamefont {M.}~\bibnamefont
  {Shen}}\ and\ \bibinfo {author} {\bibfnamefont {D.~R.}\ \bibnamefont
  {Nicholson}},\ }\href@noop {} {\bibfield  {journal} {\bibinfo  {journal}
  {Phys. Fluids},\ }\textbf {\bibinfo {volume} {30}},\ \bibinfo {pages} {1096}
  (\bibinfo {year} {1987})}\BibitemShut {NoStop}%
\bibitem [{\citenamefont {{Matthews}}\ \emph {et~al.}(1976)\citenamefont
  {{Matthews}}, \citenamefont {{Pongratz}},\ and\ \citenamefont
  {{Papadopoulos}}}]{MatthewsA_1976_JGR}%
  \BibitemOpen
  \bibfield  {author} {\bibinfo {author} {\bibfnamefont {D.~L.}\ \bibnamefont
  {{Matthews}}}, \bibinfo {author} {\bibfnamefont {M.}~\bibnamefont
  {{Pongratz}}}, \ and\ \bibinfo {author} {\bibfnamefont {K.}~\bibnamefont
  {{Papadopoulos}}},\ }\Doi {10.1029/JA081i001p00123} {\bibfield  {journal}
  {\bibinfo  {journal} {J. Geophys. Res.},\ }\textbf {\bibinfo {volume} {81}},\
  \bibinfo {pages} {123} (\bibinfo {year} {1976})}\BibitemShut {NoStop}%
\bibitem [{\citenamefont {Hughes}\ and\ \citenamefont
  {LaBelle}(2001)}]{HughesLaBelle_2001_JGR}%
  \BibitemOpen
  \bibfield  {author} {\bibinfo {author} {\bibfnamefont {J.~M.}\ \bibnamefont
  {Hughes}}\ and\ \bibinfo {author} {\bibfnamefont {J.}~\bibnamefont
  {LaBelle}},\ }\href@noop {} {\bibfield  {journal} {\bibinfo  {journal} {J.
  Geophys. Res.},\ }\textbf {\bibinfo {volume} {106}},\ \bibinfo {pages}
  {21157} (\bibinfo {year} {2001})}\BibitemShut {NoStop}%
\bibitem [{\citenamefont {{Papadopoulos}}\ and\ \citenamefont
  {{Coffey}}(1974){\natexlab{b}}}]{PapadopoulosCoffey_1974_JGR_anomalous}%
  \BibitemOpen
  \bibfield  {author} {\bibinfo {author} {\bibfnamefont {K.}~\bibnamefont
  {{Papadopoulos}}}\ and\ \bibinfo {author} {\bibfnamefont {T.}~\bibnamefont
  {{Coffey}}},\ }\Doi {10.1029/JA079i010p01558} {\bibfield  {journal} {\bibinfo
   {journal} {J. Geophys. Res.},\ }\textbf {\bibinfo {volume} {79}},\ \bibinfo
  {pages} {1558} (\bibinfo {year} {1974}{\natexlab{b}})}\BibitemShut {NoStop}%
\bibitem [{\citenamefont {{Rowland}}\ \emph
  {et~al.}(1981){\natexlab{b}}\citenamefont {{Rowland}}, \citenamefont
  {{Papadopoulos}},\ and\ \citenamefont {{Palmadesso}}}]{RowlandA_1981_GRL}%
  \BibitemOpen
  \bibfield  {author} {\bibinfo {author} {\bibfnamefont {H.~L.}\ \bibnamefont
  {{Rowland}}}, \bibinfo {author} {\bibfnamefont {K.}~\bibnamefont
  {{Papadopoulos}}}, \ and\ \bibinfo {author} {\bibfnamefont {P.~J.}\
  \bibnamefont {{Palmadesso}}},\ }\Doi {10.1029/GL008i012p01257} {\bibfield
  {journal} {\bibinfo  {journal} {Geophys. Res. Lett.},\ }\textbf {\bibinfo
  {volume} {8}},\ \bibinfo {pages} {1257} (\bibinfo {year}
  {1981}{\natexlab{b}})}\BibitemShut {NoStop}%
\end{thebibliography}
\end{document}